\documentclass[11pt,twoside]{osajnl}
\journal{josaa}
\setboolean{shortarticle}{false} 
\usepackage{amsmath,amsfonts,amssymb}
\usepackage{graphicx}
\usepackage[]{aas_macros}
\usepackage{color}

\def\bA{\boldsymbol{A}}

\def\ra{\mathrm{a}}

\def\bC{\boldsymbol{C}}
\newcommand{\rd}{\mathrm{d}} 
 
\def\bD{\boldsymbol{D}}

\def\mcE{\mathcal{E}}
\def\bF{\boldsymbol{F}}

\def\rg{\mathrm{g}}

\def\rm{\mathrm{m}}

\def\bn{\boldsymbol{n}}

\def\rp{\mathrm{p}}

\def\bQ{\boldsymbol{Q}}

\def\rr{\mathrm{r}}
\def\bR{\boldsymbol{R}}
\def\bs{\boldsymbol{s}}

\def\rT{\mathrm{T}}

\def\bU{\boldsymbol{U}}
\def\bu{\boldsymbol{u}}

\def\bV{\boldsymbol{V}}
\def\bw{\boldsymbol{w}}
\def\rw{\mathrm{w}}
\def\bW{\boldsymbol{W}}
\def\bx{\boldsymbol{x}}
\def\by{\boldsymbol{y}}
\def\bz{\boldsymbol{z}}

\def\btheta{\boldsymbol{\theta}}

\title{Treating Wavefront Measurement Error in Estimation of Non-Common Path Aberration for Direct Imaging of Exoplanets }

\author[$\dagger$*]{Richard A. Frazin}
\affil[$\dagger$]{ Dept. of Climate and Space Sciences and Engineering, University of Michigan, Ann Arbor, MI 48109} 
\affil[*]{E-mail: rfrazin@umich.edu}

\dates{Compiled \today}

\ociscodes{010.1080 Adaptive Optics, 010.7350   Wavefront Sensing}

\doi{\url{http://dx.doi.org/10.1364/ao.XX.XXXXXX}}

\begin{abstract}

One of the major difficulties limiting ground-based direct imaging of exoplanets with adaptive optics is quasi-static speckles in the science camera (SC) that obscure the planetary image.
These speckles are caused by aberrations, called non-common path aberrations (NCPA), that are not corrected in the adaptive optics loop, and all attempts to subtract them in post-processing have been problematic.
The method of Frazin (2013) (F13) uses simultaneous millisecond telemetry from wavefront sensor (WFS) and the SC to estimate the both the NCPA and the exoplanet image in a self-consistent manner.
Rodack et al. (2018) proposed correcting for the NCPA in real-time while on-sky using the F13 estimation method, and called this procedure the "Real-Time Frazin Algorithm."
The original regression model underlying the F13 method did not account for uncertainty in the WFS measurements, and this cannot be done with standard statistical methodology since these uncertainties manifest themselves in the independent variables (i.e., they cannot be treated as another source of noise in the SC data).
Further, simulations show that simply using the noisy wavefront measurements without accounting for their uncertainties leads to estimates of the NCPA with unacceptably large bias.
Here, the source of this bias is explained in terms of an "errors in variables" statistical model.
Then, the method of F13 is generalized to account for WFS measurement error using a new sequential estimation technique that treats the nonlinear coupling between NCPA, WFS measurements and the error covariance of the WFS measurements.
This new technique keeps a running estimate of the NCPA, the exoplanet image and their joint covariance matrix.
The sequential implementation of the method should make it computationally efficient enough to be suitable for on-sky correction of the NCPA as well as off-line analysis.

\end{abstract}
\setboolean{displaycopyright}{true}
\begin{document} 
\maketitle
\thispagestyle{fancy}

\section{Introduction}\label{sec: intro}

In direct imaging of exoplanets from a ground-based telescope , the most promising method for achieving the needed contrast combines high-order adaptive optics (AO), sometimes called ``extreme" AO (ExAO), with a stellar coronagraph.
A stellar coronagraph is a telescopic imaging system designed to block the light from a star on the optical axis while only having minimal effect on the portion of the image surrounding the star \cite{Sivar01}.
In ground-based observing, the efficacy of the coronagraph depends on the quality of the wavefront correction provided by the AO system.
Typically, the AO system operates at visible wavelengths, while the science camera in the coronagraph captures images in the near-infrared (IR) at wavelengths of 1.2 $\mu$m or longer.
Ideally, the phase of the wavefront entering the portion of the optical train containing the coronagraph is uniform (i.e., flat), in which case the starlight suppression is extremely efficient with any modern stellar coronagraph design \cite{Guyon_LyotPIAA14, Krist_WFIRST2016}. 
But, a combination of atmospheric turbulence and aberrations introduced by the telescope hardware require correction with an AO loop in order to make wavefront entering the corongraph as flat as possible.

The most problematic aberrations for exoplanet imaging are caused by the telescope hardware are the \emph{non-common path aberrations} (NCPA), which occur in optical components that are beyond the dichroic filter (or beam splitter) that separates the light paths that go to the AO system's wavefront sensor (WFS) and the coronagraph.
The term NCPA arises because these aberrations are not experienced by the light going to both the coronagraph and the WFS, thus, AO system is not directly controlling the wavefront going to the coronagraph.
The NCPA manifest themselves as \emph{quasi-static speckles} in the coronagraph's science camera (SC).
The term ``quasi-static" is applied because the NCPA because they evolve on time-scales ranging from minutes to hours, in contrast to the millisecond time-scales of the variation of the atmospheric aberrations, and several authors have shown that they are the major limitation in high-contrast imaging \cite{Boccaletti04, Martinez13}.
The quasi-static speckles are particularly problematic, because, unlike the speckles created by the atmospheric turbulence, they do not average to spatially smooth halo that can be subtracted in post-processing in a relatively straightforward manner.

Recently, a number of researchers have proposed various methods to differentiate planetary signal from the unwanted stellar speckles using millisecond exposure times in the SC.
Millisecond exposure are interesting because the atmospheric turbulence begins to effectively freeze at these time-scales, and the resulting SC images show a swarm of atmospheric speckles that change with each exposure.
Millisecond exposure times are becoming observationally attractive due to a new generation of noiseless IR and near-IR detectors capable of millisecond read-out times \cite{SWIR_detector14,SELEX_APD12,Saphira_eAPD14,Mazin_MKIDS14}.
Building on the pioneering work of Ref.~\cite{Gladysz10}, Frazin (2013), henceforth F13, showed analytically that an adaptive optics system will create a signal in the SC in which the planetary image and the stellar speckle behave much differently in time \cite{Frazin13}.
F13 further demonstrates that a regression procedure can take advantage of simultaneous millisecond telemetry from the WFS and the SC to make self-consistent estimate of the exoplanet image and the non-common path aberrations (NCPA). 

Of the several methods that utilize the millisecond telemetry only F13 and Ref.~[\citenum{Codona13}] also utilize the WFS telemetry.
The other methods simply discard the information from the WFS.
Additionally, F13 is the only method that estimates the NCPA that cause the quasi-static aberrations, while the others attempt to differentiate between speckles and planets (in principle, the F13 method can reconstruct the speckle and planetary image sequences after the needed regressions have been carried out).
F13's unique ability estimate the NCPA led Rodack et al. (2018), henceforth R18, to propose implementing the F13 in real-time to correct the NCPA on-sky, and they called the method the ``real-time Frazin algorithm" (RTFA) \cite{Rodack18}.
As of this article, the F13 algorithm will account for measurement error in the WFS telemetry, further differentiating the F13 method from  Ref.~[\citenum{Codona13}].

One method that potentially competes with F13, \emph{Speckle deconvolution} (SD) uses ms images and takes advantage of the different temporal behavior of the stellar speckle and the exoplanet image.
This method has the advantage that it does not require any modeling of the optical system, but it has several disadvantages relative to F13 \cite{Gladysz10}.
The disadvantages of SD relative to F13 are:
\begin{itemize}
\item{SD does not take advantage of the WFS telemetry.}
\item{The method assumes the histogram of the stellar speckle intensity can be described by a modified Rician density \cite{StatisticalOptics}.
This cannot be strictly true since the ``constant phasor'' component isn't constant, rather, it follows the instantaneous value of the Strehl ratio.  This effect probably will be detrimental when it comes to estimating the planetary intensity.}
\item{The SD method is only formulated for the point in the SC that is at the center of the planet's image.  Thus, it is valid for the pixel at the center of the planet image under the (safe) assumption that the pixel size is less than $\lambda/D$.
Thus, the method cannot use the planetary photons that arrive at pixels that are a bit further from the center of the planetary image.
In contrast, F13 makes use of all of the information.}
\end{itemize}

The F13 regression is relatively straightforward when we ignore the error in the measurement of the wavefront by the WFS, however simply ignoring this problem produces unacceptably biased estimates of the NCPA and planetary image.
Including this error significantly complicates the estimation problem.
Resolving this issue in a principled manner that is efficient enough for on-sky correction of the NCPA and determination of the planetary image is the subject of this article.

\section{Problem Statement and Optical Model}

Let us focus our attention on the problem of utilizing noisy wavefront information for estimations of the NCPA and the exoplanet image.
To get started, let the unknown NCPA that we seek to estimate be represented by the vector $\bx_\ra$, which contains the $N_\ra$ NCPA coefficients.
These could be coefficients of Zernike modes or any other representation of the aberration.
Similarly, the vector $\bx_\rp$ will contain the $N_\rp$ coefficients describing the planetary image.
These coefficients could correspond to a pixel-based description of the planetary image or some other representation such as splines (which would require a smaller number of coefficients than pixels).
The total number of coefficients that we need to estimate is $N = N_\ra + N_\rp$.
We will assume that $\bx_\rp$ is constant in time since treating temporal variation in the planetary image on the time-scale of one or several nights of observing should not be necessary in most cases and is an even more difficult problem.
The objective of the R18 paper is to estimate the NCPA and send a correction signal to the deformable mirror (DM) in real-time.
Since the small-amplitude NCPAs will not be apparent on the ms time-scales, it may take minutes or more for them to be reliable estimated from the data.
To this end, we will assume that their coefficients, $\bx_\ra$, are constant in time over $T$\ time-steps in the main AO loop.
$T$\ will also be called the ``batch size," which will correspond to the inverse frequency of the NCPA correction loop.
The strategy we will adopt here is to assume that $\bx_\ra$\ changes from one batch to the next but is constant within a batch.
On the other hand, we need to build up our knowledge of $\bx_\ra$\ over many batches, so what we have learned about $\bx_\ra$ must be passed from batch to batch.
The majority of the discussion will treat the single-batch problem and later I will show how the knowledge $\bx_\ra$ can be transferred in a straightforward fashion.

For simplicity, we will assume that the SC and WFS have synchronous exposures of roughly 1 ms in duration, indexed by the time-step $t$.
At time $t$, the parameters that specify the wavefront entering the WFS are contained in the vector $\bw_t$; it is the job of the WFS to estimate $\bw_t$.
For example, $\bw_t$ could contain pixel-by-pixel phase values, or it could contain coefficients of Zernike modes the specify the phase.
There is no reason that $\bw_t$\ cannot contain amplitude information as well.
Now, the wavefront entering the coronagraph will be different than the one measured by the WFS due to the NCPA, so, the stellar intensity in the SC at time $t$ will be a function of both $\bw_t$\ and $\bx_\ra$.
On the other hand, I will assume that the NCPA are small enough have no effect on to the planetary image, so that the planetary contribution to the intensity in the SC is independent of the $\bx_\ra$.
F13 requires an optical model of the coronagraph that predicts the intensity in the SC as a function of $\bw_t$, $\bx_\ra$ and $\bx_\rp$.
While this model is linear $\bx_\rp$, it is nonlinear in $\bw_t$\ and $\bx_\ra$ \cite{Frazin13}.
Then, the vector of instantaneous SC intensities at time $t$, represented by the $M \times 1$\ vector $\by_t$, can be expressed in terms of the full nonlinear model of the coronagraph as:
\begin{equation}
\by_t = \bA_\ra(\bw_t, \bx_\ra) + \bA_\rp(\bw_t)\bx_\rp + \bn_t
\label{eq: SC model at t}
\end{equation}
where $\bA_\ra(\bw_t, \bx_\ra)$\ is a $M \times N_\ra$ matrix of functions that model stellar component of the SC intensity, $\bA_\rp$\ is planetary optical model, which is represented as a $M \times N_p$\ matrix of functions, and $\bn_t$\ represents noise in the SC. 
At first glance, calculating the models $\bA_\ra$ and $\bA_\rp$ is likely to be computationally intensive, involving large Fourier transforms and the like.
However, the same method that accelerates pyramid wavefront sensor computations in Ref.~\cite{Frazin_JOSAA2018} can be applied here.
In the method of Ref.~\cite{Frazin_JOSAA2018}, one pre-computes the electric field in the detector plane that results from set of point-sources in the entrance aperture.
Once these fields have been calculated, then electric field in the detector plane that results from any other desired input wavefront can be determined with simple matrix-vector multiplication.
Thus, after an initial set of expensive calculations have been performed, the remaining calculations are quite easy and rapid.

In the statistical inference procedures described below, the following definitions will help to economize notation::
\begin{eqnarray}
\bx & \equiv & \left[ \begin{array}{c} \bx_\ra \\ \bx_\rp \end{array} \right] 
 , \: \: \mathrm{and}
 \label{eq: x def} \\
\bA(\bw_t, \bx) & \equiv &   \bA_\ra(\bw_t, \bx_\ra) + \bA_\rp(\bw_t)\bx_\rp    \, .
\label{eq: A def}
\end{eqnarray}
Thus, the $\bx$ vector that contains both the planetary image and the NCPA coefficients, and $\bA(\bw_t, \bx)$\ combines the stellar and planetary optical models.

Since we expect the NCPA to be small, linearizing $\bA_\ra$ with respect $\bx_\ra$ is a good way to start this analysis.
Including the full nonlinearity should not be much more difficult using methods similar to those I applied in Ref.~\cite{Frazin_JOSAA2018} to the pyramid wavefront sensing problem.
In F13, the coronagraph model included 2\underline{nd} order (quadratic) terms in $\bx_\ra$, although there is an important algebra error in accounting for these terms that was pointed out in the later published erratum.
As an aside, I point out that treating the nonlinearity via a method similar to one employed in Ref.~\cite{Frazin_JOSAA2018} is probably a better strategy than dealing with the quadratic terms directly as in F13.
Linearizing $\bA_\ra$ with respect $\bx_\ra$ yields:
\begin{eqnarray}
\bA_\ra(\bw_t, \bx_\ra) & \approx & \bu(\bw_t) + \bA_\ra(\bw_t) \bx_\ra \: \: ,  \: \mathrm{where}
\label{eq: linearized A_a at t} \\
&& \: \: \: \: \:  \bu(\bw_t) \equiv \bA_\ra(\bw_t, 0) \,
\label{eq: u def}
\end{eqnarray}
where $\bA_\ra(\bw_t)$\ is a $M \times N_\ra$\ Jacobian matrix:
\begin{equation}
\bA_\ra(\bw_t) \equiv \frac{\partial \bA_\ra(\bw_t, \bx_\ra)}{\partial \bx_\ra} \bigg|_0
\label{eq: A Jacobian}
\end{equation}
In \eqref{eq: linearized A_a at t}, $\bu(\bw_t)$\ is the vector of the stellar component SC intensities predicted by the optical model when the wavefront is given by $\bw_t$ in the absence of NCAP (i.e., $\bx_\ra = 0$).
One way to interpret \eqref{eq: linearized A_a at t} is that in the absence of the NCPA and planetary light, with a wavefront specified by $\bw_t$, $\bu(\bw_t)$ would be the prediction that our optical model makes for the SC intensity values.
Combining \eqref{eq: linearized A_a at t} with \eqref{eq: SC model at t}, we arrive at:
\begin{equation}
\by_t = \bu(\bw_t) + \bA_\ra(\bw_t) \bx_\ra + \bA_\rp(\bw_t)\bx_\rp + \bn_t  \,
\label{eq: linearized SC model at t, full expression}
\end{equation}
which accounts for small NCPA and the planetary contribution in the SC.
\eqref{eq: linearized SC model at t, full expression} can be written more compactly as:
\begin{equation}
\by_t  = \bu(\bw_t) + \bA(\bw_t) \bx + \bn_t \, ,
\label{eq: linearized SC model at t}
\end{equation}  
where
\begin{equation}
\bA(\bw_t) \equiv \big[ \bA_\ra(\bw_t) \: \: \bA_\rp] \, .
\label{eq: A partition on x}
\end{equation}

\section{Bayesian Treatment of Linear Systems with Gaussian Noise}\label{sec: Standard Treatment}

In order to illustrate several elementary concepts in Bayesian regression, I begin with the canonical framework for the linear least-squares regression problem.\cite{Moon&Stirling}
Ref.~[\citenum{Barrett07}] treated maximum-likelihood estimation in wavefront sensing, and Ref.~[\citenum{Demoment89}] provides an excellent review of Bayesian estimation in image processing.
The equations in this section are in no way specialized to exoplanet imaging, but the various quantities are defined in a way that is consistent with the rest of the discussion.
Let me first introduce convenient notation for the multivariate normal distribution over the random vector variable $\bs$, which as $M$\ components, centered on $\bs_0$, with $M \times M$ (positive definite) covariance matrix $\bC$:
\begin{equation}
\mathcal{N} \big( \bs ; \bs_0 , \bC  \big) \equiv
\frac{1}{\sqrt{(2\pi)^M |\bC|}} \exp \left[- \frac{1}{2} (\bs - \bs_0)^\rT \bC^{-1} (\bs - \bs_0) \right] \, ,
\label{eq: normal notation}
\end{equation}
in which $^\rT$\ indicates transpose, $|\bC|$ is the determinant of $\bC$, and $\bs_0$\ and $\bC$\ are taken to be constants.

The linear regression problem can be stated as:
\begin{equation}
\by = \bA \bx + \bn \, ,
\label{eq: canonical linear}
\end{equation}
where $\bA$ is an $M \times N$ matrix of \emph{independent variables}, and the vectors $\by$, $\bn$ and $\bx$ are $M \times 1$, $M \times 1$ and $N \times 1$, respectively.
In \eqref{eq: canonical linear}, $\bA$\ is assumed to known exactly, $\by$\ represents measured values (which are known once the measurements have been made), $\bn$\ represents noise in the measurements and is thus unknown, but is assumed to be governed by a zero-mean normal probability density with known $M \times M$ covariance matrix $\bC$, which must be positive definite.
In this linear regression problem, we seek to estimate the \emph{``state of the system'' (or simply ``state'')} $\bx$\ given the values of $\bA$, $\by$ and $\bC$.
The \emph{measurement model} expresses the probability (before the measurements are made) of measuring the values $\by$ assuming that the value of the state is $\bx$ and is represented formally as the \emph{conditional} probability density $P(\by | \bx)$:
\begin{align}
P(\by | \bx) & = \frac{1}{\sqrt{(2\pi)^M |\bC|}} \exp \left[- \frac{1}{2} (\by - \bA \bx)^\rT \bC^{-1} (\by - \bA \bx) \right] \nonumber \\
& = \mathcal{N}(\by; \bA \bx, \bC)
\label{eq: Gaussian linear measurement}
\end{align}
which is also called the \emph{likelihood}.

To get the \emph{maximum likelihood} (ML) estimate of $\bx$, we maximize $ - \ln [ P(\by | \bx)]$ from \eqref{eq: Gaussian linear measurement} with respect to $\bx$.
It is easy to show this results in:
\begin{equation}
\hat{\bx}_\mathrm{ML} = (\bA^\rT \bC^{-1} \bA)^{-1}\bA^\rT \bC^{-1} \by \, ,
\label{eq: max likelihood}
\end{equation}  
which is the usual formula for weighted least-squares.
Now, if $\bA^\rT \bC^{-1} \bA$ is singular, then it cannot be inverted and $\hat{\bx}_\mathrm{ML}$ in \eqref{eq: max likelihood} does not exist.
Even if $\bA^\rT \bC^{-1} \bA$ is non-singular but has one or more small eigenvalues (making it a poorly conditioned matrix), then $\hat{\bx}_\mathrm{ML}$ will often be unacceptable due to amplification of noise.
In these cases it is necessary to apply some form of regularization.
Within a Bayesian context, this is done with so-called \emph{prior} probability density on the state $P(\bx)$.
Formally, the prior $P(\bx)$ can be viewed as a \emph{marginal} of the joint density $P(\bx, \by)$, but in most cases of practical interest the form $P(\bx)$ is simply chosen for the sake of mathematical convenience not because it actually represents the marginalization of the joint density (which is often not available).
For example, a common choice of $P(\bx)$ is:
\begin{equation}
P(\bx) = \mathcal{N}[\bx; \bx_0, (\gamma \bR)^{-1}  ] \propto
\exp \left( - \frac{\gamma}{2} (\bx - \bx_0)^\rT \bR (\bx - \bx_0) \right) \, ,
\label{eq: regularization prior}
\end{equation}
where the vector $\bx_0$\ is introduced as the mean of the prior, and the normalization is of no concern here.
In  \eqref{eq: regularization prior}, the $N \times N$\ matrix $\bR$ is called the \emph{regularization matrix} and $\gamma \geq 0$\ is the regularization parameter.
Common choices of $\bR$\ include $\mathbb{1}$ (the identity matrix) and $\bD^\rT \bD$, where $\bD$ is a finite-difference matrix.
These choices enforce smoothness on the solution.
Note that it is permissible for these purposes for $\bR$\ to be singular, so long as the matrix $(\bA^\rT \bC^{-1} \bA + \gamma \bR)$ is not singular.
If $\bR$ is singular note that $P(\bx)$ is not a proper probability density (since its integral over $\bx$ would diverge). 
Since $P(\by, \bx) = P(\by | \bx) P(\bx)$, we can combine Eqs.~(\ref{eq: Gaussian linear measurement}) and (\ref{eq: regularization prior}) to get something formally equivalent to the joint density function:
\begin{equation}
P(\by, \bx) = P(\by | \bx) P(\bx) \propto
\exp \left[- \frac{1}{2} (\by - \bA \bx)^\rT \bC^{-1} (\by - \bA \bx) - \frac{\gamma}{2} (\bx - \bx_0)^\rT \bR (\bx - \bx_0) \right] \, .
\label{eq: joint regularized}
\end{equation}
The marginal density $P(\by)$ can be obtained by integrating \eqref{eq: joint regularized} over $\bx$ for the linear Gaussian model.

The \emph{posterior} density is defined as the conditional density $P(\bx | \by)$, i.e., the probability of the state being $\bx$ given the measurements $\by$.
From a Bayesian point-of-view,2 the posterior density carries all of the available knowledge about $\bx$ once the measurements $\by$ have been collected.
Bayes' rule relates the joint density to the conditionals: $P(\bx, \by) \, = \, P(\by | \bx) P(\bx) \,  = \, P(\bx | \by) P(\by)$, so that the posterior density can be expressed as $P(\bx | \by) = P(\by | \bx) P(\bx) / P(\by)$.
The \emph{maximum a-posteriori} (MAP) estimate of $\bx$ is defined as the value of $\bx$ that maximizes the posterior probabilty:
\begin{equation}
\hat{\bx}_\mathrm{MAP}  = \underset{\bx}{\mathrm{argmax}} \left[ \frac{P(\by | \bx) P(\bx)}{P(\by)}   \right]
 = \underset{\bx}{\mathrm{argmax}} \left[ P(\by | \bx) P(\bx)   \right] \, ,
\end{equation}
where the second equality follows because the denominator does not depend on $\bx$. 
Therefore, for the linear Gaussian model, the MAP estimate is found by maximizing the probability in \eqref{eq: joint regularized} with respect to $\bx$.
It is easy to show that the result is:
\begin{equation}
\hat{\bx}_\mathrm{MAP} = (\bA^\rT \bC^{-1} \bA + \gamma \bR)^{-1} \big[ \bA^\rT \bC^{-1} \by + \gamma \bR \bx_0 \big] \, ,
\label{eq: MAP}
\end{equation}
where it is assumed that the matrix $(\bA^\rT \bC^{-1} \bA + \gamma \bR^\rT \bR)$ is non-singular, which it will be for any reasonable choice of $\bR$.
From \eqref{eq: MAP}, one can see that as the regularization parameter $\gamma$ increases, the solution becomes more regularized (smoother).

Let us now consider several expectations using the previous developments.
For a fixed state $\bx$, the expected value of $\by$, otherwise known as the \emph{conditional expectation}, written as $\mcE (\by | \bx)$ is defined as:
\begin{equation}
\mcE(\by | \bx) \equiv \int \rd \by \, \by P(\by | \bx)   \, .
\label{eq: def expected y}
\end{equation}
Using \eqref{eq: Gaussian linear measurement} in \eqref{eq: def expected y}, it is straightforward to show:
\begin{equation}
\mcE (\by | \bx) = \bA \bx  \, .
\label{eq: def condition mean y}
\end{equation}
The \emph{conditional covariance} matrix of $\by$ is defined similarly to \eqref{eq: def expected y}:
\begin{align}
\mcE\{ [\by - \mcE(\by|\bx)][ \by - \mcE(\by|\bx) ]^\rT | \bx \} & \equiv 
\int \rd \by  \, [\by - \mcE(\by|\bx)][ \by - \mcE(\by|\bx) ]^\rT P(\by | \bx)  \\
& = \bC \, ,
\label{eq: def conditional covar y}
\end{align}
where we have first used \eqref{eq: def condition mean y} to replace $\mcE(\by|\bx)$ with $\bA \bx$ and then applied \eqref{eq: Gaussian linear measurement} when performing the integration.
This rather obvious ``result'' can be seen as the definition of the covariance matrix $\bC$.
Below, we will require an expression for $\mcE( \by \by^\rT | \bx)$, which, using  \eqref{eq: Gaussian linear measurement}, is:
\begin{equation}
\mcE( \by \by^\rT | \bx) = \bC + \bA\bx \bx^\rT \bA^\rT \, .
\label{eq: mean y y^T}
\end{equation}

Now using Eqs.~(\ref{eq: max likelihood}) and~(\ref{eq: def condition mean y}), we have:
\begin{eqnarray}
\mcE(\hat{\bx}_\mathrm{ML} | \bx) & = & (\bA^\rT \bC^{-1} \bA)^{-1}\bA^\rT \bC^{-1} \mcE (\by | \bx) \nonumber \\
& = & \bx \,   
\label{eq: unbiased MLE}
\end{eqnarray}
Thus, the ML estimate is \emph{unbiased} (assuming it exists).
The conditional covariance matrix of the ML estimator is defined as:
\begin{align}
\mathrm{covar}(\hat{\bx}_\mathrm{ML} | \bx)  \equiv \:  & 
\mcE \left\{ \big[ \hat{\bx}_\mathrm{ML} - \mcE(\hat{\bx}_\mathrm{ML} | \bx)   \big]
\big[ \hat{\bx}_\mathrm{ML} - \mcE(\hat{\bx}_\mathrm{ML} | \bx)   \big]^\rT | \bx \right\} \nonumber \\
 = \: & 
\mcE \left\{ \big[ \hat{\bx}_\mathrm{ML} - \bx    \big]  \big[ \hat{\bx}_\mathrm{ML} - \bx    \big]^\rT
  | \bx \right\} \nonumber \\
 = \: & \mcE( \hat{\bx}_\mathrm{ML}  \hat{\bx}^\rT_\mathrm{ML} | \bx)  - \bx \bx^\rT   \nonumber \\
 = \: & (\bA^\rT \bC^{-1} \bA)^{-1}\bA^\rT \bC^{-1} \mcE (\by \by^\rT | \bx) [(\bA^\rT \bC^{-1} \bA)^{-1}\bA^\rT \bC^{-1}]^\rT  - \bx \bx^\rT  \nonumber \\ 
 = \:  & (\bA^\rT \bC^{-1} \bA)^{-1} \, ,
\label{eq: covar MLE}
\end{align}
where we have made use of \eqref{eq: mean y y^T} and the fact that the inverse of a symmetric matrix is symmetric [e.g., $(\bC^{-1})^\rT = \bC^{-1}$]. 
Now, if we calculate the mean and covariance of the MAP estimators with respect to the same conditional density $P(\by | \bx)$ in \eqref{eq: Gaussian linear measurement}, we find:
\begin{equation}
\mcE(\hat{\bx}_\mathrm{MAP} | \bx) = (\bA^\rT \bC^{-1} \bA + \gamma \bR)^{-1}\bA^\rT \bC^{-1} \big[ \bA \bx + \gamma \bR \bx_0 \big]
\label{eq: biased MAP}
\end{equation}
Since the matrices $ \bA^\rT \bC^{-1} \bA $ and $  \gamma \bR$ are both positive semi-definite, 
the matrix $(\bA^\rT \bC^{-1} \bA + \gamma \bR)$\ will have larger eigenvalues than $\bA^\rT \bC^{-1} \bA$, so that $| \mcE(\hat{\bx}_\mathrm{MAP} | \bx) | \leq | \bx |$ if $\bx_0 = 0$.
Thus, the MAP estimate is biased even when $\bx_0 = 0$.
Indeed, this bias is a desired property since the solution is pushed towards being smooth in the way specified by the form of the regularization matrix $\bR$.
Following similar methodology, we can calculate the variances of the ML and MAP estimators.
The covariance matrix of the MAP estimator is approximately given by:
\begin{align}
\mathrm{covar}(\hat{\bx}_\mathrm{MAP} | \bx) & =  (\bA^\rT \bC^{-1} \bA + \gamma \bR)^{-1} (\bA^\rT \bC^{-1} \bA ) (\bA^\rT \bC^{-1} \bA + \gamma \bR)^{-1}
\label{eq: covar MAP exact} \\ 
& \approx \; (\bA^\rT \bC^{-1} \bA + \gamma \bR)^{-1}  \, ,
\label{eq: covar MAP}
\end{align}
where we have used the approximation $(\bA^\rT \bC^{-1} \bA + \gamma  \bR)^{-1} (\bA^\rT \bC^{-1} \bA) \approx \mathbb{1}$, which is valid if $\gamma$ is small enough (assuming $\bA^\rT \bC^{-1}\bA$ is non-singular).
If $ (\bA^\rT \bC^{-1} \bA) $ is non-singular, then we have$ |\mathrm{var}(\hat{\bx}_\mathrm{MAP} ) | \leq | \mathrm{var}(\hat{\bx}_\mathrm{ML})  |$ due to the positive semi-definiteness of  the matrices $ \bA^\rT \bC^{-1} \bA $ and $  \gamma \bR$.

\section{Errors in Variables for the Coronagraph System}\label{Errors in Variables}

In Sec.~\ref{sec: Standard Treatment} we established a practical and widely used framework for least-squares regression of linear equations, which treats errors in the measurements $\by$ in a rather straightforward fashion.
The problem becomes much more difficult if we must consider imperfect knowledge of the matrix $\bA$.
This is called the \emph{errors in variables problem}, and it continues to be an active field in statistics research \cite{Stefanski_nonlinear}.
Wikipedia has some introductory material in the article entitled ``Errors-in-variables models."
Errors in variables arise in our context due to the fact that we have an imperfect measurement of the phase and amplitude of the wavefront that is seen by the wavefront sensor, which is required as an input to the F13 method.

To simplify our discussion we will work in the small aberration regime, so that the model is linear in $\bx$, as shown in \eqref{eq: linearized SC model at t}.
Let us now concatenate $T$ time-steps together to make a version of \eqref{eq: linearized SC model at t} that describes the concatenated SC data, which will be represented by the vector $\by \equiv [\by^\rT_0, \, \cdots, \, \by^\rT_{T-1}]^\rT$ (the transpose operations are required so that the result is a long column vector).
Applying the same stacking procedure to $\bA(\bw_t)$\ and $\bn_t$ in \eqref{eq: linearized SC model at t}, it becomes:
\begin{equation}
\by  = \bu(\bw) + \bA(\bw) \bx + \bn \, ,
\label{eq: linearized SC model}
\end{equation}  
where the time-series of wavefronts is given by $\bw \equiv [\bw^\rT_0, \, \cdots, \, \bw^\rT_{T-1}]^\rT$, and similarly for $\bA$ and $\bn$.

Clearly, if time-series of wavefronts ($\bw$) were known exactly, we could subtract $\bu(\bw)$\ from both sides of  \eqref{eq: linearized SC model} and apply the the standard ML or MAP estimation procedures described above to make an estimate of $\bx$.
As things stand, \eqref{eq: linearized SC model} is essentially a measurement model, but now it is conditioned on both $\bw$\ and $\bx$.
I.e., assuming that the noise ($\bn$) is zero-mean and can be approximated by a normal density with covariance matrix $\bC_{\by} \,$, the measurement model in \eqref{eq: Gaussian linear measurement} takes the form:
\begin{equation}
P(\by | \bw, \bx) = \frac{1}{\sqrt{(2\pi)^{MT} |\bC_{\by}|}} \exp \left\{- \frac{1}{2} \big[\by - \bu(\bw) - \bA(\bw) \bx \big]^\rT 
\bC_{\by}^{-1} \big[\by - \bu(\bw) - \bA(\bw) \bx \big] \right\} \, .
\label{eq: complicated Gaussian linear measurement}
\end{equation}

\subsection{Naive Estimation} 

The most simple way to incorporate the measurement of $\bw$\ from the WFS, which will denote as $\hat{\bw}$, is to replace $\bw$\ in \eqref{eq: complicated Gaussian linear measurement} with $\hat{\bw}$, which results in an expression called the \emph{pseudo-likelihood}:
\begin{equation}
P_\mathrm{naive}(\by | \bx) = \frac{1}{\sqrt{(2\pi)^{MT} |\bC_{\by}|}} \exp \left\{- \frac{1}{2} \big[\by - \bu(\hat{\bw}) - \bA(\hat{\bw}) \bx \big]^\rT 
\bC_{\by}^{-1} \big[\by - \bu(\hat{\bw}) - \bA(\hat{\bw}) \bx \big] \right\} \, .
\label{eq: pseudo-likelihood}
\end{equation}
If we find the value of $\bx$\ that maximizes the pseudo-likelihood in \eqref{eq: pseudo-likelihood}, we arrive at the so-called \emph{naive} estimate of $\bx$:
\begin{equation}
\hat{\bx}_\mathrm{ML \; naive} = \big[ \bA^\rT(\hat{\bw}) \bC_{\by}^{-1} \bA(\hat{\bw}) \big]^{-1}
\bA^\rT(\hat{\bw}) \bC_{\by}^{-1} \big[ \by - \bu(\hat{\bw}) \big]    \, ,
\label{eq: ML naive estimate}
\end{equation}
which has the same mathematical form as the maximum likelihood estimate in \eqref{eq: max likelihood}.
Naive estimation can take advantage of a prior $P(\bx) $\ as in \eqref{eq: joint regularized} to get a naive MAP estimate:
\begin{equation}
\hat{\bx}_\mathrm{MAP \; naive} = \big[ \bA^\rT(\hat{\bw}) \bC_{\by}^{-1} \bA(\hat{\bw}) + \gamma \bR^\rT \bR) \big]^{-1}\bA^\rT(\hat{\bw}) \bC_{\by}^{-1} \big[ \by - \bu(\hat{\bw}) \big]    \, ,
\label{eq: MAP naive estimate}
\end{equation}
which is in the same form as \eqref{eq: MAP}.
While the naive estimate does correctly account for the noise in $\by$, it does not account for the error in $\hat{\bw}$.
Even if  $\hat{\bw}$ is an unbiased estimate $\bw$ and $\bA(\bw)$\ is linear in $\bw$, the naive estimator in \eqref{eq: ML naive estimate} is nonlinear in $\hat{\bw}$\ and is likely to have significant bias, as is explained in a large body of statistical literature.
My simulations and those of A. Rodack confirm this for our problem.

In order to gain some mathematical understanding of the origin of this bias, let us represent the error in the wavefront measurement by $\delta \bw$\ defined as:
\begin{equation}
\delta \bw \equiv \hat{\bw} - \bw \, ,
\label{eq: error w}
\end{equation}
where the sign is chosen for later convenience.
Similarly, the resulting errors in optical model are given by
\begin{align}
\delta \bA(\bw) \equiv & \: \bA(\hat{\bw}) - \bA(\bw)  \: \: \:  \mathrm{and}
\label{eq: error A} \\
\delta \bu(\bw) \equiv & \: \bu(\hat{\bw}) - \bu(\bw) \,  .
\label{eq: error u}
\end{align}
Using Eqs.~(\ref{eq: error A}) and (\ref{eq: error u}), \eqref{eq: ML naive estimate} can be written as:
\begin{align}
\hat{\bx}_\mathrm{ML \; naive} = & \left\{ [\bA^\rT(\bw) + \delta \bA^\rT(\bw) ]\bC_{\by}^{-1} [ \bA(\bw) + \delta \bA(\bw)] \right\}^{-1} 
\nonumber \\
& \: \: \: \times [\bA^\rT(\bw) + \delta \bA^\rT(\bw)]\bC_{\by}^{-1} \big[ \by - \bu(\bw) - \delta \bu(\bw) \big] \, .
\label{eq: delta naive MLE} 
\end{align}
Now, the inverse in \eqref{eq: delta naive MLE} can be approximated with a series expansion in matrix powers of $\delta \bA(\bw)$.
Clearly, the expression that results from such an expansion will have terms that are 1\underline{st}, 2\underline{nd} and higher order in $\delta \bA(\bw)$ and of $\delta \bu(\bw)$.
Even if expectations  $\delta \bA(\bw)$ and $\delta \bu(\bw)$\ with respect to the error statistics of $\hat{\bw}$ are zero, the expansion terms of 2\underline{nd} order and higher will have have non-zero expectations, which are a source of bias in the naive estimate.
That said, the optical model itself $\bA(\bw)$ is nonlinear in $\bw$ (see F13), so even the 1\underline{st} order terms can contribute bias.

\subsection{Simultaneous Estimation}\label{Simultaneous Estimation}

In this section I introduce a strategy to alleviate the bias issue inherent in naive estimation. 
The basic idea is to find expression for the joint likelihood of $\bx$\ and $\bw$ and maximize this joint likelihood with respect to $\bx$ and $\bw$\ simultaneously.  
By maximizing the joint likelihood, the estimate of $\bx$\ is informed by the probability density that governs the estimate of $\bw$, so the effects of bias are mitigated.
The likelihood of the SC data at time is given by the nonlinear measurement model:
\begin{equation}
P(\by_t | \bw_t, \bx ) = \frac{1}{\sqrt{(2\pi)^M |\bC_{\by , t}|}} \exp 
 \left\{ - \frac{1}{2} \big[\by_t  - \bA(\bw_t, \bx) \big]^\rT  \bC_{\by, t}^{-1} \big[\by_t - \bA(\bw_t , \bx) \big] \right\}
\label{eq: SC measurement}
\end{equation}
where the vector $\by_t$ is the set of SC intensities values for the the current time $t$ and $\bC_{\by,t}$ is the measurement noise covariance for $\by_t$.
Let me now introduce the WFS measurements at time $t$ in the form of the vector $\bz_t$\ in which $\bz_t$\ are the (zero-point subtracted) intensities in the WFS camera at time $t$.
Assuming that these measurements are provided by a pyramid wavefront sensor (PyWFS), we can use a model of the PyWFS optical system as shown in Ref.~\cite{Frazin_JOSAA2018} to create the nonlinear measurement model:
\begin{equation}
P(\bz_t | \bw_t) = \frac{1}{\sqrt{(2\pi)^L |\bC_{\bz , t}|}} \exp \left\{- \frac{1}{2} [\bz_t - \bW(\bw_t)]^\rT \bC_{\bz, t}^{-1} [\bz_t - \bW (\bw_t) ] \right\} \, ,
\label{eq: WFS measurement}
\end{equation}
where the vector of WFS measurement $\bz_t$ has $L$\ components with covariance matrix $\bC_{\bz ,t }$\ and $\bW(\bw_t)$\ is the nonlinear optical system model of the WFS.
Ref.~\cite{Frazin_JOSAA2018} shows how ML and MAP estimation can be applied to both linearized and nonlinear PyWFS measurements in order to get estimates of $\bw_t$.
Assuming the densities in Eqs.~(\ref{eq: SC measurement}) and (\ref{eq: WFS measurement}) are statistically independent, they can be combined into a joint likelihood at time $t$ as:
\begin{equation}
 P(\by_t , \bz_t | \bw_t \, , \bx ) = P(\bz_t | \bw_t) P(\by_t | \bw_t, \bx ) 
\label{eq: joint measurement at t}
\end{equation}
Assuming that the densities at the time-steps are statistically independent, the joint likelihood of over all of the time-steps can be written as:
\begin{equation}
P(\bz, \by | \bw, \bx) =  P(\bz | \bw) P(\by | \bw, \bx) = \prod_{t=0}^{T-1}  P(\bz_t | \bw_t) P(\by_t | \bw_t, \bx ) \, ,
\label{eq: grand joint likelihood}
\end{equation}
in which the concatenated wavefront camera intensities are given by: $\bz = [\bz^\rT_0, \, \cdots, \, \bz^\rT_{T-1}]^\rT$.
Similarly to \eqref{eq: joint regularized}, we can now write the full joint density of $\bz, \, \by, \, \bw$ and $\bx$ by multiplying the joint likelihood in \eqref{eq: grand joint likelihood} by the prior densities $P(\bx)$ and $P(\bw) = P(\bw_0) \cdots P(\bw_{T-1})$:
\begin{equation}
P(\bz, \by , \bw, \bx) = P(\bx) \prod_{t=0}^{T-1} P(\bw_t)  P(\bz_t | \bw_t) P(\by_t | \bw_t, \bx ) \, .
\label{eq: full joint factorization} 
\end{equation}
To avoid the biases inherent in naive estimation, we maximize the density in \eqref{eq: full joint factorization} with respect to $\bx$ and $\bw_t$ simultaneously, to yield the pair of estimates $(\hat{\bx}, \hat{\hat{\bw}})$, i.e.,
\begin{equation}
(\hat{\bx}, \hat{\hat{\bw}}) = \underset{(\bx, \bw)}{\mathrm{argmax}} \big[ P(\bz, \by , \bw, \bx) \big]   ,
\label{eq: full joint MAP}
\end{equation}
which is a MAP estimate.
Note that the double hat notation ($\hat{\hat{\bw}}$) is used to distinguish the estimate of $\bw$\ in \eqref{eq: full joint MAP} from $\hat{\bw}$, which is the set of estimates of the wavefronts $\bw$\ that may be derived from analysis of the WFS data alone [i.e., \eqref{eq: WFS measurement}], which are denoted as $\hat{\bw} = [\hat{\bw}^\rT_0, \, \cdots, \, \hat{\bw}^\rT_{T-1}]^\rT$.

In principle, performing the optimization problem in \eqref{eq: full joint MAP} solves our problem, since that solution fully accounts for the uncertainty in the WFS and SC measurements.
The key feature that distinguishes simultaneous estimation of $\bx$ and $\bw$ via \eqref{eq: full joint MAP} from naive estimation is the nonlinearity in the $\bA(\bw_t, \bx)$ term that couples $\bx$ and $\bw_t$, which would require an iterative optimization method such as conjugate gradient or a Newton-type scheme.  
Note that this nonlinearity will be important even in the limit that $\bx \rightarrow 0$, so as long as the likelihood governing $\bw_t$ in \eqref{eq: WFS measurement} does not constrain $\hat{\bw_t}$\ enough to make the error terms in the naive estimate in \eqref{eq: delta naive MLE} negligible. 
Numerically, the various nonlinearities in log of the probability in \eqref{eq: full joint factorization} should not be too problematic since the pair $(\hat{\bx}_\mathrm{naive} , \, \hat{\bw})$ should be an excellent initial guess for the optimization.
The main problem with maximizing the probability in \eqref{eq: full joint MAP} is that with a one millisecond exposure time (and extreme-AO systems run even more quickly) $T \sim 6 \times 10^4$ for a 1 minute batch of data, so $\bw = [\bw_0^\rT, \, \cdots, \bw_{T-1}^\rT]^\rT$ is a very large vector indeed.
In the next section we will see how sequential estimation should make optimization problem in \eqref{eq: full joint MAP} computationally tractable.

\section{Sequential Estimation}\label{sec: Sequential Estimation}

Sequential estimation is often used to analyze time-series data due to the computational efficiency of various algorithms, of which  Kalman filtering is the most widely known \cite{Anderson&Moore}.
I will avoid a standard introduction to Kalman filterning here, except to say that it can be viewed as an extension of MAP estimation to time-dependent quantities.
There is a large literature on Kalman filtering and related methods, so here I will optimize the discussion to support an approach that I think will best serve purpose of maximizing the probability in \eqref{eq: full joint factorization} for our problem.

Let us consider the problem of ingesting the SC data at time-step $t$.
By ``ingesting" we mean using the SC measurements at made time-step $t$ to update our knowledge of $\bx$.
In this case, ``knowledge" corresponds to the mean and covariance of the probability density that governs our estimate of it.
As in the Kalman filter, our knowledge of $\bx$\ after ingesting the information at $t-1$\ is used as a prior, i.e.:
\begin{equation}
P_{\bx , t}(\bx | \hat{\bx}_{t-1} , \bC_{\hat{\bx},  t-1}  ) = \mathcal{N}\big( \bx; \, \hat{\bx}_ {t-1} , \bC_{\hat{\bx}, t-1} \big) \, , 
\label{eq: x t-1 prior}
\end{equation} 
where $\hat{\bx}_{t-1}$\ is the value of the MAP estimate of $\bx$ from the previous time-step and $\bC_{\hat{\bx}, t-1}$ is its covariance matrix, both of which are assumed to be known, so that \eqref{eq: x t-1 prior} is a density over one random (vector) variable $\bx$.
As in \eqref{eq: full joint factorization}, we will continue to assume $P(\bw_t)$ is prior distribution of $\bw_t$, so that we can use \eqref{eq: x t-1 prior} to write an expression for a conditional joint probability:
\begin{equation}
P(\bz_t, \by_t, \bw_t, \bx | \hat{\bx}_{t-1} , \bC_{\hat{\bx},  t-1} ) =
 P(\bw_t) \mathcal{N}\big( \bx; \, \hat{\bx}_ {t-1} , \bC_{\hat{\bx}, t-1} \big) P(\bz_t | \bw_t) P(\by_t | \bw_t, \bx) \,
\label{eq: conditional joint}
\end{equation} 
where the final two factors are the likelihoods given in Eqs.~(\ref{eq: WFS measurement}) and (\ref{eq: SC measurement}).
Once $\bz_t$ and $\by_t$ have been measured by the WFS and SC, respectively, \eqref{eq: conditional joint} is a probability density over two random vectors $\bw_t$ and $\bx$.
Then, the sequential analog of \eqref{eq: full joint MAP} is:
\begin{equation}
(\hat{\bx}, \hat{\hat{\bw}}_t) = \underset{(\bx, \bw_t)}{\mathrm{argmax}} \big[ P(\bz_t, \by_t, \bw_t, \bx | \hat{\bx}_{t-1} , \bC_{\hat{\bx},  t-1} )  \big]   ,
\label{eq: sequential joint MAP}
\end{equation}
\eqref{eq: sequential joint MAP} is in a form that allows fusion of the SC and WFS data to make an optimal estimate of $\bw_t$.
Such fusion estimates of $\bw_t$ are potentially useful when the WFS data alone are insufficiently informative to measure the desired aspects of the wavefront, as happens with the ``island effect'' and phasing of segmented mirrors \cite{Schwartz_IslandEffect2018}.
In addition, when used with no modulation, the pyramid wavefront sensor is a highly nonlinear device \cite{Frazin_JOSAA2018, Hutterer_nonlinearPyWFS2018}, and the combined SC and WFS data would likely help to mitigate this problem.

The optimization problem in \eqref{eq: sequential joint MAP} cannot be avoided if the WFS data alone are not sufficient to make a useful estimate of $\bw_t$, which is denoted as $\hat{\bw}_t$.
However, if we have $\hat{\bw}_t$ and its error covariance matrix $\bC_{\hat{\bw},t}$  available from analysis of the WFS signal, then in \eqref{eq: conditional joint},  we can make the replacement:
\begin{equation}
P(\bw_t)  P(\bz_t | \bw_t) \rightarrow \mathcal{N} \big( \bw_t ; \hat{\bw}_t , \bC_{\hat{\bw}, t} \big) \, .
\label{eq: w_t replacement}
\end{equation}
$\mathcal{N} \big( \bw_t ; \hat{\bw}_t , \bC_{\hat{\bw}, t} \big)$ is not a function of $\bz_t$, but contains all of our knowledge about $\bw_t$\ after processing the measurements $\bz_t$.
Note that in Eqs.~(\ref{eq: conditional joint}) and (\ref{eq: w_t replacement}) $P(\bw_t)$\ is considered to be an uninformative prior used for regularization.
Then, instead of \eqref{eq: conditional joint}, we have:
\begin{equation}
P(\by_t, \bw_t, \bx | \hat{\bx}_{t-1} , \bC_{\hat{\bx},  t-1} ) =
\mathcal{N} \big( \bw_t ; \hat{\bw}_t , \bC_{\hat{\bw}, t} \big)
 \mathcal{N}\big( \bx; \, \hat{\bx}_ {t-1} , \bC_{\hat{\bx}, t-1} \big)  P(\by_t | \bw_t, \bx) \, .
\label{eq: alt conditional joint}
\end{equation} 
Then, that the pair $(\hat{x}, \hat{\hat{\bw}}_t)$ is instead estimated via:
\begin{equation}
(\hat{\bx}, \hat{\hat{\bw}}_t) = \underset{(\bx, \bw_t)}{\mathrm{argmax}} \big[ P(\by_t, \bw_t, \bx | \hat{\bx}_{t-1} , \bC_{\hat{\bx},  t-1} )  \big]   .
\label{eq: alt sequential joint MAP}
\end{equation}
The advantage of \eqref{eq: alt sequential joint MAP} over \eqref{eq: sequential joint MAP} is that $\mathcal{N} \big( \bw_t ; \hat{\bw}_t , \bC_{\hat{\bw}, t} \big)$\ will have a much simpler functional form than does $P(\bz_t | \bw_t) $, leading to an easier (and faster) optimization problem.
The optimization problems in either \eqref{eq: sequential joint MAP} and \eqref{eq: alt sequential joint MAP} require vastly less computation than does \eqref{eq: full joint MAP} since only a single time-step is being processed instead of the full batch of $T$ time-steps.
Even so, the optimization still must confront the $\bA(\bw_t , \bx)$ nonlinearity that couples $\bw_t$ and $\bx$ in \eqref{eq: SC measurement}, even if the other nonlinearities are not significant.
Note that due to the nonlinearities, there are possibilities of arriving at a local minimum that is not the global minimum.
Any nonlinear optimization algorithm will be iterative and require a starting guess.
A good initial guess for the iterative optimization is the pair $(\bx_\mathrm{g}, \hat{\bw}_t)$,  where $\hat{\bw_t}$\ is some estimate of $\bw_t$ made from WFS data alone.
A guess value $\bx_\rg $, can be found by using the linearization of the optical model in \eqref{eq: linearized SC model at t} in the measurement model $P(\by_t | \hat{\bw}_t, \bx)$\ from \eqref{eq: SC measurement}, and maximizing $ \mathcal{N}\big( \bx; \, \hat{\bx}_ {t-1} , \bC_{\hat{\bx}, t-1} \big)  P(\by_t | \hat{\bw}_t, \bx) $ over $\bx$ analytically:
\begin{align}
\hat{\bx}_\rg \; & = \; \underset{\bx}{\mathrm{argmax}} \big[  \mathcal{N}\big( \bx; \, \hat{\bx}_ {t-1} , \bC_{\hat{\bx}, t-1} \big)  P(\by_t | \hat{\bw}_t, \bx)   \big]  \nonumber \\
   & \approx  \; \big[  \bA( \hat{\bw}_t )^\rT \bC_{\by, t}^{-1} \bA(\hat{\bw}_t) \, + \, \bC_{\hat{\bx}, t-1}^{-1} \big]^{-1} \nonumber  \\
& \: \: \: \: \: \: \: \times \left\{  \big[  \bA( \hat{\bw}_t )^\rT \bC_{\by, t}^{-1}  \big] \big[ \by_t - \bu( \hat{\bw}_t  )  \big]
  +  \bC_{\hat{\bx},t-1}^{-1} \hat{\bx}_{t-1}  \right\} \, ,
\label{eq: x_g}
\end{align}
which is essentially a naive estimate in the same form as \eqref{eq: MAP}, and where the approximation is due to the linearization from \eqref{eq: linearized SC model at t}.

\subsection{Finding the Covariance of the Estimate}

So far, we have seen how to efficiently ingest the data at time $t$, assuming that we have access $\hat{\bx}_{t-1}$ and $\bC_{\hat{\bx}, t-1}$.
While  $\hat{\bx}_{t-1}$ comes directly from either \eqref{eq: sequential joint MAP} or \eqref{eq: alt sequential joint MAP} at the previous time-step, calculating $\bC_{\hat{\bx}, t}$ is more challenging, and it requires calculation of an approximate error-covariance matrix of the joint estimate $(\hat{\bx}_t, \hat{\hat{\bw}}_t)$.
For this, we first need to derive the Fisher information matrix, \cite{Moon&Stirling, Barrett07}, the inverse of which establishes a lower bound the variance of any unbiased estimator.
I expect that the estimation procedure, i.e. \eqref{eq: sequential joint MAP} or (\ref{eq: alt sequential joint MAP}), will be nearly unbiased, but that should be verified with detailed simulation.

Before calculating the Fisher infomation, we must first decide which probability density model governs the pair $(\bx, \bw_t)$, which will be either \eqref{eq: sequential joint MAP} or \eqref{eq: alt sequential joint MAP}.
Let $f(\bx, \bw_t)$ be the one we choose, i.e.,
\begin{align}
f(\bx, \bw_t) & \; \equiv \;  P(\bz_t, \by_t, \bw_t, \bx | \hat{\bx}_{t-1} , \bC_{\hat{\bx},  t-1} ) \nonumber \\
& \: \: \: \: \: \: - \: \mathrm{or} \: -
\label{eq: density choice} \\
& \; \equiv \;  P(\by_t, \bw_t, \bx | \hat{\bx}_{t-1} , \bC_{\hat{\bx},  t-1} ) \, . \nonumber 
\end{align}
This compact notation emphasizes that the density $f$ is over the random vector variables $\bx$ and $\bw_t$, while the other quantities present in $P(\bz_t, \by_t, \bw_t, \bx | \hat{\bx}_{t-1} , \bC_{\hat{\bx},  t-1})$ or  $P(\by_t, \bw_t, \bx | \hat{\bx}_{t-1} , \bC_{\hat{\bx},  t-1} ) $, given in \eqref{eq: conditional joint} or (\ref{eq: alt conditional joint}), are constants. 

In the notation common to many discussions of multivariate estimation theory, the vector to be estimated is represented by the symbol $\btheta$.
Since we are estimating the two vectors, $\bx$\ and $\bw_t$, the corresponding $\btheta$ is given by the concatenation $\btheta = [\bx^\rT , \bw_t^\rT]^\rT$. and we define $f(\btheta) \equiv f(\bx, \bw_t)$
The Fisher information is then defined by the 2\underline{nd} derivative matrix:
\begin{eqnarray}
\bF & = &- \mcE \left[ \frac{\partial^2}{\partial \btheta \partial \btheta} \ln f(\btheta)   \right]
\label{eq: compact Fish} \\
& = & - \mcE
\left[ \begin{array}{c c}
  \frac{\partial^2}{\partial \bx \partial \bx}  \ln f(\bx, \bw_t)  &  \frac{\partial^2}{\partial \bx \partial \bw_t} \ln f(\bx, \bw_t) \\
\left(\frac{\partial^2}{\partial \bx \partial \bw_t} \ln f(\bx, \bw_t) \right)^\rT &  \frac{\partial^2}{\partial \bw_t \partial \bw_t} \ln f(\bx, \bw_t) 
\end{array}\right]   \, ,
\label{eq: big Fish}
\end{eqnarray}
which is composed of the smaller Fisher information matrices associated with $\bx$, $\bw_t$ as well as the one that mixes $\bx$ and $\bw_t$.
The expectations are with respect the density $f$, i.e., formally, the expectation involves integration over the random variables $\bx$ and $\bw_t$.
While evaluating the expectation of some function of some probability density with respect to that same probability density is counter-intuitive, that is what is required.

It is instructive to demonstrate computation of the Fisher information.
First, we must go to \eqref{eq: density choice} in order to choose the probability density model.
For simplicity, let us take the second option, i.e., $f(\bx, \bw_t) \equiv P(\by_t, \bw_t, \bx | \hat{\bx}_{t-1} , \bC_{\hat{\bx},  t-1} ) $.
Using \eqref{eq: alt conditional joint}, we find:
\begin{align}
- 2 \ln f(\bx, \bw_t) = \, & c \, + \,  [\by_t - \bA(\bw_t, \bx)]^\rT \bC_{\by, t} [\by_t - \bA(\bw_t, \bx)] \nonumber \\
& +  (\hat{\bx}_{t-1} - \bx)^\rT \bC_{\hat{\bx}, t-1}^{-1} (\hat{\bx}_{t-1} - \bx)
  + (\hat{\bw}_t - \bw_t)^\rT \bC_{\hat{\bw}, t}^{-1} (\hat{\bw}_t - \bw_t)
\label{eq: log f}
\end{align}
where $c$\ is a normalization constant that can be determined from \eqref{eq: alt conditional joint}.
To proceed further, it is helpful to linearize the function $\bA(\bw_t, \bx)$ about the new estimate pair $(\hat{\bx}_t, \hat{\hat{\bw}}_t)$ in both $\bx$ and $\bw_t$, so that:
\begin{equation}
\bA(\bw_t, \bx) \approx \bA(\hat{\hat{\bw}}_t, \hat{\bx}_t) + \bU (\bx - \hat{\bx}_t) + \bV(\bw_t - \hat{\hat{\bw_t}}) \,
\label{eq: A linearized}
\end{equation}
where $\bU$\ and $\bV$\ are the Jacobian matrices:
\begin{align}
\bU \equiv  &  \; \frac{\partial \bA(\bw_t, \bx)}{\partial \bx}\bigg|_{\hat{\hat{\bw}}, \hat{\bx}_t}
\label{eq: U def} \\
\bV \equiv &  \;  \frac{\partial \bA(\bw_t, \bx)}{\partial \bw_t}\bigg|_{\hat{\hat{\bw}}, \hat{\bx}_t}
\label{eq: V def}  
\end{align}
Note that the linearization in \eqref{eq: A linearized} differs from the one in  \eqref{eq: linearized SC model at t} since we are linearizing about a different point and we are linearizing with respect to both $\bw_t$ and $\bx$.
Also, the matrices $\bU$\ and $\bV$ both have $M$ rows (the number of pixels in the SC), but they will probably differ in their number of columns since the vectors $\bw_t$ and $\bx$ will likely have different lengths.

The upper left block of the matrix in \eqref{eq: big Fish} is the $(\bx, \bx)$\ 2\underline{nd} derivative matrix.
Using the linearization in \eqref{eq: A linearized}, it is easy to show that the elements of this block are given by:
\begin{eqnarray}
\mcE \left[ \frac{- \partial^2}{\partial \bx \partial \bx}  \ln f(\bx, \bw_t) \right]
& \approx & \mcE  \big[ \bC_{\hat{\bx}, t-1}^{-1} + \bU^\rT \bC_{\by, t}^{-1} \bU \big] \, , \nonumber \\
& = & \bC_{\hat{\bx}, t-1}^{-1} + \bU^\rT \bC_{\by, t}^{-1} \bU \, ,
\label{eq: linearized Fisher x-x}
\end{eqnarray}
where the approximation is due to the linearization and the final equality follows since the expectation of a constant  matrix is itself.
The other two blocks in the Fisher information matrix in \eqref{eq: big Fish} can be approximated in a similar manner.

With the Fisher information matrix approximated, then we can bound the error covariance matrix of the pair $(\hat{\bx}_t, \hat{\hat{\bw}}_t)$ with the Cramer-Rao inequality \cite{Moon&Stirling}:
\begin{equation}
\mathrm{covar}\big( \hat{\bx}_t, \hat{\hat{\bw}}_t   \big) \geq \mathcal{F}^{-1}  \, .
\end{equation}
Since this procedure only provides a lower bound on the error covariance of the estimate, numerical simulations will be needed to help determine whether or not this bound is an adequate approximation.
If the inverse of the Fisher information matrix is found to be a problematic underestimate of the estimate error covariance, then we will need to some inflation strategy.

\subsection{Batch Initialization}\label{sec: Batch Init}
So far, we have seen how to proceed from time-step $t$ to $t+1$ within a batch, but what about starting a new batch?
Since the batch starts with $t=0$, we need to initialize the estimation sequence with $\hat{\bx}_{-1}$\ and the matrix $\bC_{\hat{\bx},-1}$.
The way we initialize these quantities will allow us to carry or discard information from one batch to the next.
Recalling that $\bx = [\bx_\ra^\rT, \, \bx_\rp^\rT]^\rT$, the error covariance matrix $\bC_{\hat{\bx}, \, t}$ can be partitioned into four blocks as:
\begin{equation}
\bC_{\hat{\bx}, \, t} =
\left[ \begin{array}{c c}
\bC_{\ra , \ra, t} & \bC_{\ra, \rp, t } \\
 \bC_{\ra, \rp, t }^\rT & \bC_{\rp, \rp, t}
\end{array}\right] \, ,
\label{eq: x partition}
\end{equation}
which is valid for any time-step $t$ (starting with $-1$ and ending with $T-1$), and where  $\bC_{\ra , \ra, t}$ and $ \bC_{\rp, \rp, t}$ correspond to the error covariance matrices of $\hat{\bx}_{\ra, t}$ and $\hat{\bx}_{\rp, t}$, respectively, and $\bC_{\rp , \ra, t}$\ is the cross-covariance matrix.
The partition in \eqref{eq: x partition} provides a guide for the initialization procedure.
At the very beginning of observing (i.e., starting the 0\underline{th} batch) nothing is known about the planetary image or the NCPA, so we can set $\bx_{-1} = 0$\ and $\bC_{\hat{\bx}, \, -1}$\ to a diagonal matrix with the largest values that numerical stability will allow.

The initialization between batches requires more careful consideration so as not to discard what has learned about $\bx_\rp$\ and $\bx_\ra$ from previous batches.
Our knowledge of the planetary image itself is preserved if we simply make the transfer
\begin{eqnarray}
\hat{\bx}_{\rp, T-1} & \rightarrow & \hat{\bx}_{\rp, -1} \, , \: \mathrm{and} \label{eq: init x_p} \\
\bC_{\rp, \rp, T-1} & \rightarrow  & \bC_{\rp, \rp, -1} \label{eq: init C_pp}
\end{eqnarray}
Now, the entire point of the RTFA is to correct for the NCPA after they have been estimated from a batch.
Let $\hat{\bx}_{\ra, T-1}$\ be the estimate of the NCPA portion of $\hat{\bx}_{T-1}$ from the previous batch, and let $\hat{\delta \bx}_\ra$\ be the estimated value of the correction applied (since it will not be known exactly) in the loop running the RTFA.
Then, $\hat{\bx}_{\ra, -1}$ should be initialized to the current best estimate, i.e.,:
\begin{equation}
\hat{\bx}_{\ra, T-1} - \hat{\delta \bx}_\ra \, \rightarrow \hat{\bx}_{\ra, -1} \, .
\label{eq: init x_a} 
\end{equation} 
The matrices $\bC_{\ra , \ra, t} $ and $\bC_{\ra, \rp, -1 }$ need to be initialized as well.
I suggest the initializations:
\begin{eqnarray}
\bC_{\ra , \ra, T-1} + \bQ_{\ra , \ra} & \rightarrow & \bC_{\ra , \ra, -1}  \, , \: \mathrm{and} \label{eq: init C_aa} \\
\bC_{\ra , \rp, T-1} + \bQ_{\ra , \rp} & \rightarrow & \bC_{\ra , \rp, -1} \, , \label{eq: init C_ap}
\end{eqnarray}
where $\bQ_{\ra , \ra}$ $\bQ_{\ra , \rp}$ are an \emph{inflation matrices} chosen to partially erase the memory of the NCPA estimate from the previous batch.
In this way, the estimate of the NCPA can evolve from batch-to-batch with only the desired amount of memory.  
The simplest choice for $\bQ_{\ra , \ra}$ is a diagonal matrix, but any positive definite matrix is mathematically admissible.
If $\bQ_{\ra , \ra}$ is too small, then it will be hard for the estimate of the NCPA to follow the temporal evolution of the true NCPA.
On the other hand, if it is too large, then it will be hard to find smaller amplitude NCPA that can only be inferred from multi-batch time-series.

The simplest choice for  $\bQ_{\ra , \rp}$ is 0, since it is unclear how best to inflate the cross-covariance matrix.
Mathematically, $\bQ_{\ra , \rp}$ must be chosen so that the full covariance matrix $ \bC_{\hat{\bx},  -1}$ is positive definite.
Since the sum of two positive definite matrices is another positive definite matrix,  $ \bC_{\hat{\bx},  -1}$  guaranteed to be positive definite if the matrix 
\begin{equation}
\bQ =
\left[ \begin{array}{c c}
\bQ_{\ra, \ra} & \bQ_{\ra, \rp} \\
 \bQ_{\ra, \rp}^\rT & 0
\end{array}\right] \, ,
\label{eq: Q partition}
\end{equation}
is positive semi-definite (it cannot be positive definite due to the block of zeros on the diagonal).

In summary, the RTFA implemented in batches via sequential estimation, for time-steps indexed by $t$ with $t \in [0, \, \dots , \, T-1]$ is:
\begin{enumerate}
\item{If t=0: Initialize the vector $\hat{\bx}_{-1}$ and the covariance matrix $\bC_{\hat{\bx},-1}$\ as explained in Sec.~\ref{sec: Sequential Estimation}.\ref{sec: Batch Init}.}
\item{Given $\by_t \, , \; \hat{\bx}_{t-1} \, , \; \bC_{\hat{\bx}_{t-1}} \, , \; \hat{\bw}_t \, , \; \, \bC_{\hat{\bw}, t}$,  set up the non-linear optimization corresponding to the joint probability density function in \eqref{eq: conditional joint} or \eqref{eq: alt conditional joint}.  }
\item{Maximize the probability in  \eqref{eq: conditional joint} or \eqref{eq: alt conditional joint} to yield the pair $(\hat{\bx}_t, \hat{\hat{\bw}}_t)$\ and calculate $\bC_{\hat{\bx}_{t-1}}$ via the Fisher information matrix.}
\item{If $t=T-1$: Calculate and implement the NCPA correction, $\hat{\delta \bx_\ra}$.  Start a new batch by re-initializing: $t \rightarrow 0$.  Go to step 1.}
\item{If $t < T-1$: Set $t \rightarrow t+1$. Go to Step 2 to process the next time-step.}
\end{enumerate}

\section{Model Evaluation}

At the foundation of the method proposed in this article is the optical model of the coronagraph in \eqref{eq: SC model at t}.
There are three aspects to the model that are most critical:
\begin{itemize}
\item{the base corongraph model assuming no NCPA, i.e., $\bA_\ra(\bw_t, 0) = \bu(\bw_t)$}
\item{the parameterization of the NCPA in terms of the vector $\bx_\ra$, i.e., ``what do these coefficients represent?''}
\item{how the model treats the NCPA, i.e., $\bA_\ra(\bw_t, \bx_\ra) - \bA_\ra(\bw_t, 0) \,$.}
\end{itemize}
In contrast, parameterizing the planetary image in terms of the coefficients $\bx_\rp$\ and calculating the planetary image in the SC via the perhaps simplified optical model $\bA_\rp(\bw_t)\bx_\ra$ should be relatively straightforward and will not be discussed further.
It will undoubtedly be necessary to evaluate the how well these choices represent the optical system.
To that end, once $\hat{\bx}_k$\ (where $k \in [0, \, \dots , \, T-1]$) and $\hat{\hat{\bw}}_t$\ have been estimated, the predicted SC image at time $t$ is given by:
\begin{equation}
\by_t' = \bA \big( \hat{\hat{\bw}}_t, \, \hat{\bx}_k \big) \, , 
\label{eq: predicted intensity} 
\end{equation}
the most useful value of $k$ is probably the final one, i.e., $k=T-1$.

One common way to evaluate model quality is with a goodness-of-fit measured known as the \emph{reduced} $\chi^2$ \emph{metric}.
Generally speaking, if $\chi^2_\rr >> 1$ (where $\chi^2_\rr$ is the reduced $\chi^2$), the model is considered to be poor.
Perhaps the model is not working working due to too few free parameters, but it could be the result of a more fundamental problem.
The \emph{over-fitting} regime corresponds to $\chi^2_\rr << 1$, and the model is thought too many free parameters (or the covariance matrix over-estimated).  
If we are working with the joint density in \eqref{eq: alt conditional joint}, the reduced $\chi^2$\ will be given by:
\begin{align}
\chi^2_\rr = &
\frac{1}{T(M + 2N_\rw) + N  } \sum_{t=0}^{T-1} \bigg[ (\by_t' - \by_t)^\rT \bC_{\by, t}^{-1} (\by_t' - \by_t) \nonumber \\
& \: + \: (\hat{\hat{\bw}}_t - \hat{\bw}_t)^\rT \bC_{\hat{\bw},t}^{-1} (\hat{\hat{\bw}}_t - \hat{\bw}_t) \bigg] \, ,
\label{eq: chi-squared}
\end{align}
where $N_\rw$\ is the number of components in the $\bw_t$ vector.
If we work with \eqref{eq: conditional joint} instead of \eqref{eq: alt conditional joint}, we need to change the $\chi^2$ expression accordingly. 
In \eqref{eq: chi-squared} has a sum over time-steps indexed by $t$, and each time step has the contribution from the SC and WFS measurements.
The $\chi_\rr^2$\ metric must include these two components because both $\{ \by_t \}$\ and $\{ \hat{\bw}_t \}$ play the role of observations in the likelihood in \eqref{eq: alt conditional joint}.
The normalization of $\chi_\rr^2$ is the number of degrees of freedom, which is total number of measurements plus the number free parameters.
In this case, there are $T$ time-steps and, at each time-step, there are $M$ SC measurements plus $N_\rw$ WFS measurements.
In addition, there are $N$\ free parameters in $\hat{\bx}$ and $N_\rw$ free parameters in $\hat{\hat{\bw}}_t$ at each time-step.
In \eqref{eq: chi-squared}, one can see that as $\hat{\hat{\bw}}_t$\ is tuned to reduce the discrepancy between $\by_t'$ and $\by_t$, the value of the first term decreases, but at the expense of increasing the discrepancy between $\hat{\hat{\bw}}_t$ and $\hat{\bw}_t$ and the value of the second term.

\subsection{Model Improvement}

Assuming that the base coronagraph model $\bA_\ra(\bw_t, 0)$ is adequate, the main avenue for improving the model is the treatment of the NCPA.   
Here, we will assume that the optical model operates on the electric field in a pre-coronagraph pupil plane, and then propagates that field through coronagraph and onto the SC, and then calculates the SC intensity from the field at the SC. 
F13 treats the NCPA as a (perhaps complex-valued) phase screen in the pre-coronagraph pupil plane, but this is an assumption that requires justification.
While planar defects on a surface in or near a pupil plane in the optical system can be treated this way, other types of defects cannot.
For example, if the light is brought to a focus at a glass surface, surface errors on the glass will essentially be operating on the Fourier transform of the beam.
When the beam is later re-collimated these focal-plane defects will act a convolution with some point-spread function (PSF) on the field in the pupil plane. 
The parameters specifying this type of NCPA would then need to model this effect.
Similarly, defects in a plane that is intermediate to a pupil plane and focal plane will have a pupil plane manifestation in the form of a PSF that depends on position in the pupil plane.

Given the incentive to have a model that is as simple as possible that describes the data, it is useful to have values in the vector $\bx_\ra$ represent the parameters in a hierarchy of models in order of increasing complexity, so that the complexity and the number of parameters in $\bx_\ra$ can be increased as needed.
I suggest the following hierarchy of NCPA models, all of which operate in the pupil-plane field before applying the coronagraph model:
\begin{itemize}
\item{real-valued pupil-plane phase screen}
\item{complex-valued pupil-plane phase screeen (to model amplitude effects)}
\item{convolution kernel corresponding to a real-valued phase screen in a focal plane}
\item{convolution kernel correspond to a complex-valued phase screen in a focal plane (again, to model amplitude effects)}
\item{Fresnel propagation based model to treat a real-valued phase screen in an intermediate plane}
\item{Fresnel propagation based model to treat a complex-valued phase screen in an intermediate plane}
\end{itemize}

\section{Conclusions}

In principle, the real-time Frazin algorithm is the only method that allows estimating the non-common path aberrations and correcting them on-sky while maintaining a running estimate of the exoplanet image.
In addition, it is one of only two methods for treating quasi-static speckles that makes use of the telemetry from the WFS.
A model-based approach to wavefront sensing, such as proposed in Ref.~[\citenum{Frazin_JOSAA2018}], paired with a calibrated computational model of the WFS, is particularly suited to the needs of the F13 algorithm. 
The principle limitation in the original F13 formulation was that it did not account for error in the measurements made by the WFS, and later simulations showed that this is not acceptable.
This article provides an analysis of the issue in terms of an errors-in-variables regression model, and provides a computationally efficient solution suitable for real-time computation that is grounded in statistical principles.

One problem that users of the F13 method will confront is how best to represent the NCPA.
This article shows how to apply a $\chi^2$ metric to evaluate the model quality in the context of the new sequential processing method presented here.
In addition, a hierarchy of NCPA models in order of increasing complexity was proposed in order to help users find the most appropriate model.

\section*{Acknowledgments}
Richard Frazin would like to thank:  Len Stefanski for helping me understand errors in variables problems, Johann Gagnon Bartsch for checking the statistical content of this paper, and Alex Rodack and Laurent Jolissaint for commenting on the manuscript.
This work has been supported by NSF Award \#1600138 to the University of Michigan.


\end{document}